\documentclass[reprint,superscriptaddress,amsmath,amssymb,prc,floatfix]{revtex4-2}

\usepackage{graphicx}
\usepackage{dcolumn}
\usepackage{bm}
\usepackage{hyperref}
\usepackage[mathlines]{lineno}
\usepackage{siunitx}
\DeclareSIUnit\clight{\text{\ensuremath{c}}}
\usepackage{nicefrac}
\usepackage{placeins}
\usepackage{xcolor}

\begin{document}

\preprint{}

\title{Measurement of Spin Density Matrix Elements in $\Lambda(1520)$\\ Photoproduction at \SIrange{8.2}{8.8}{\GeV}}

\affiliation{Arizona State University, Tempe, Arizona 85287, USA}
\affiliation{National and Kapodistrian University of Athens, 15771 Athens, Greece}
\affiliation{Carnegie Mellon University, Pittsburgh, Pennsylvania 15213, USA}
\affiliation{The Catholic University of America, Washington, D.C. 20064, USA}
\affiliation{University of Connecticut, Storrs, Connecticut 06269, USA}
\affiliation{Duke University, Durham, North Carolina 27708, USA}
\affiliation{Florida International University, Miami, Florida 33199, USA}
\affiliation{Florida State University, Tallahassee, Florida 32306, USA}
\affiliation{The George Washington University, Washington, D.C. 20052, USA}
\affiliation{University of Glasgow, Glasgow G12 8QQ, United Kingdom}
\affiliation{GSI Helmholtzzentrum f\"{u}r Schwerionenforschung GmbH, D-64291 Darmstadt, Germany}
\affiliation{Institute of High Energy Physics, Beijing 100049, People's Republic of China}
\affiliation{Indiana University, Bloomington, Indiana 47405, USA}
\affiliation{Alikhanov Institute for Theoretical and Experimental Physics NRC Kurchatov Institute, Moscow 117218, Russia}
\affiliation{IKP, Forschungszentrum J\"{u}lich, D-52428 J\"{u}lich GmbH, Germany}
\affiliation{Lamar University, Beaumont, Texas 77710, USA}
\affiliation{University of Massachusetts, Amherst, Massachusetts 01003, USA}
\affiliation{Massachusetts Institute of Technology, Cambridge, Massachusetts 02139, USA}
\affiliation{National Research Nuclear University Moscow Engineering Physics Institute, Moscow 115409, Russia}
\affiliation{Norfolk State University, Norfolk, Virginia 23504, USA}
\affiliation{North Carolina A\&T State University, Greensboro, North Carolina 27411, USA}
\affiliation{University of North Carolina at Wilmington, Wilmington, North Carolina 28403, USA}
\affiliation{Old Dominion University, Norfolk, Virginia 23529, USA}
\affiliation{University of Regina, Regina, Saskatchewan S4S 0A2, Canada}
\affiliation{Universidad T\'ecnica Federico Santa Mar\'ia, Casilla 110-V Valpara\'iso, Chile}
\affiliation{Thomas Jefferson National Accelerator Facility, Newport News, Virginia 23606, USA}
\affiliation{Tomsk State University, Tomsk Polytechnic University, 634050 Tomsk, Russia;  634050 Tomsk, Russia}
\affiliation{Union College, Schenectady, New York 12308, USA}
\affiliation{Washington \& Jefferson College, Washington, Pennsylvania 15301, USA}
\affiliation{William \& Mary, Williamsburg, Virginia 23185, USA}
\affiliation{Wuhan University, Wuhan, Hubei 430072, People's Republic of China}
\affiliation{A. I. Alikhanian National Science Laboratory (Yerevan Physics Institute), 0036 Yerevan, Armenia}
\author{S.~Adhikari} \affiliation{Old Dominion University, Norfolk, Virginia 23529, USA}
\author{C.~S.~Akondi} \affiliation{Florida State University, Tallahassee, Florida 32306, USA}
\author{M.~Albrecht} \affiliation{Indiana University, Bloomington, Indiana 47405, USA}
\author{A.~Ali} \affiliation{GSI Helmholtzzentrum f\"{u}r Schwerionenforschung GmbH, D-64291 Darmstadt, Germany}
\author{M.~Amaryan} \affiliation{Old Dominion University, Norfolk, Virginia 23529, USA}
\author{A.~Asaturyan} \affiliation{A. I. Alikhanian National Science Laboratory (Yerevan Physics Institute), 0036 Yerevan, Armenia}
\author{A.~Austregesilo} \affiliation{Thomas Jefferson National Accelerator Facility, Newport News, Virginia 23606, USA}
\author{Z.~Baldwin} \affiliation{Carnegie Mellon University, Pittsburgh, Pennsylvania 15213, USA}
\author{F.~Barbosa} \affiliation{Thomas Jefferson National Accelerator Facility, Newport News, Virginia 23606, USA}
\author{J.~Barlow} \affiliation{Florida State University, Tallahassee, Florida 32306, USA}
% ...
\collaboration{The \textsc{GlueX} Collaboration}

\author{D.~I.~Glazier}
\affiliation{University of Glasgow, Glasgow G12 8QQ, United Kingdom}
\author{V.~Mathieu}
    \affiliation{\ucm}
    \affiliation{\ub}

\newcommand{\ucm}{Departamento de F\'isica Te\'orica, Universidad Complutense de Madrid and IPARCOS, 28040 Madrid, Spain}
\newcommand{\ub}{Departament de F\'isica Qu\`antica i Astrof\'isica and Institut de Ci\`encies del Cosmos, Universitat de Barcelona, Mart\'i i Franqu\`es 1, E08028, Spain}

\date{July 26, 2021}

\begin{abstract}
We report on the measurement of spin density matrix elements of the $\Lambda(1520)$ in the photoproduction reaction $\gamma p\rightarrow \Lambda(1520)K^+$, via its subsequent decay to $K^{-}p$.  The measurement was performed as part of the GlueX experimental program in Hall D at Jefferson Lab using a linearly polarized photon beam with $E_\gamma =$ \SIrange[range-phrase=--]{8.2}{8.8}{\GeV}. These are the first such measurements in this photon energy range. Results are presented in bins of momentum transfer squared, $-(t-t_\text{0})$. We compare the results with a Reggeon exchange model and determine that natural exchange amplitudes are dominant in $\Lambda(1520)$ photoproduction.
\begin{description}
\item[Key words]
$\Lambda(1520)$; SDME; MCMC; GlueX
\end{description}
\end{abstract}

\maketitle

\section{Introduction}
The GlueX experiment is dedicated to expanding our knowledge of hadrons by measuring observables for a wide variety of states. The measurement presented here contributes to this effort by studying the photoproduction process of the $\Lambda(1520)$ hyperon ($J^{P}=\nicefrac{3}{2}^-$), specifically the measurement of spin density matrix elements (SDME). SDMEs parameterize the spin polarization of a produced state and are directly related to the underlying helicity amplitudes of the production process. As such they provide tests for scattering theory models which are needed in the search for new states in the hadron spectrum, especially in the search for small signals as are expected for exotic mesons. Schilling~\textit{et~al.}~\cite{Schilling1970} showed how SDMEs of vector mesons can be directly measured via the angular distribution of their decay products, and here we extend this technique for decays of spin-\nicefrac{3}{2} states. In addition to allowing us to measure the SDMEs of a strange baryon, the $\Lambda(1520)$ is experimentally attractive because it is a relatively isolated and narrow resonance with a width of \SI{16}{\MeV/\clight^2}~\cite{Zyla:2020zbs}.\par
The $\Lambda(1520)$ was discovered in 1962 using a $K^-$ beam on a proton target~\cite{Ferro-Luzzi1962}, but since then only a few photoproduction measurements have been performed, with the majority of these at lower photon energies than the results reported in this paper. The only measurements performed in an energy range similar to that of GlueX are the differential cross sections from SLAC in 1971~\cite{Boyarski1971} using an unpolarized photon beam. In 1980 the LAMP2 experiment extracted three independent SDMEs using an unpolarized photon beam with energy between \SI{2.4}{\GeV} and \SI{4.8}{\GeV}, in addition to differential cross sections~\cite{Barber1980}. The results indicated that the production does not proceed via simple $K$ exchange. More recently, measurements at lower photon energies were published, mostly of cross sections~\cite{Muramatsu2009,Kohri2010,Wieland2011,Moriya2013,Shrestha2021}. \par
Several attempts were made to describe the photoproduction of $\Lambda(1520)$ theoretically~\cite{Nam2005,Nam2010,He2014,Yu2017}. In general, the models used a Reggeon exchange approach to describe the $t$\nobreakdash-channel production, which is expected to dominate beyond the $s$\nobreakdash-channel resonance region. Since most of the available data cover a much lower energy range than that presented here, the models are not optimized for the GlueX energy range. Yu and Kong~\cite{Yu2017}, however, used the low energy results from LAMP2~\cite{Barber1980} and high energy results from SLAC~\cite{Boyarski1971} to interpolate between available data and provide predictions for seven SDMEs in the GlueX energy range. In their model, they describe the production process in terms of $K$, $K^*$, and $K_2^*$ exchanges, together with a proton pole in the $s$\nobreakdash-channel and a contact term to preserve gauge invariance. They found that, especially at high energies, the $K_2^*$ exchange is crucial to describe the data. Since the more recent cross section data from CLAS~\cite{Moriya2013} disagrees with the LAMP2~\cite{Barber1980} data by a factor of up to two at low energies (see Fig.~4 in Ref.~\cite{Yu2017}), Yu and Kong also made predictions from the same model based on the CLAS~\cite{Moriya2013} and LEPS~\cite{Muramatsu2009,Kohri2010} data at lower energies. Both predictions will be used later to compare to our new data. Precise measurement of polarized SDMEs, such as those presented here, provide strong constraints on the production mechanisms used in models of $\Lambda(1520)$ photoproduction and will therefore help with our general understanding of photoproduction processes. \par
The structure of the paper is as follows. Section~\ref{sec:formalism} introduces the SDMEs and gives the fit function used to extract them. Section~\ref{sec:experiment} gives an overview of the experimental setup used for data taking. The event selection is presented in Section~\ref{sec:selection} and Section~\ref{sec:MCMC} covers the methods used to extract the SDMEs from a sample of $\Lambda(1520)$ events. The results are discussed in Section~\ref{sec:results}.

\section{Formalism}\label{sec:formalism}
In order to study photoproduction of the $\Lambda(1520)$, we choose to reconstruct it in its decay to $K^-p$, which has a $22.5\%$ branching fraction~\cite{Zyla:2020zbs}. Therefore, we study the reaction $\gamma p\rightarrow K^+\Lambda(1520)\rightarrow K^+K^-p$.\par
\begin{figure}[tpb]
    \begin{center}
        \includegraphics[width=0.9\linewidth]{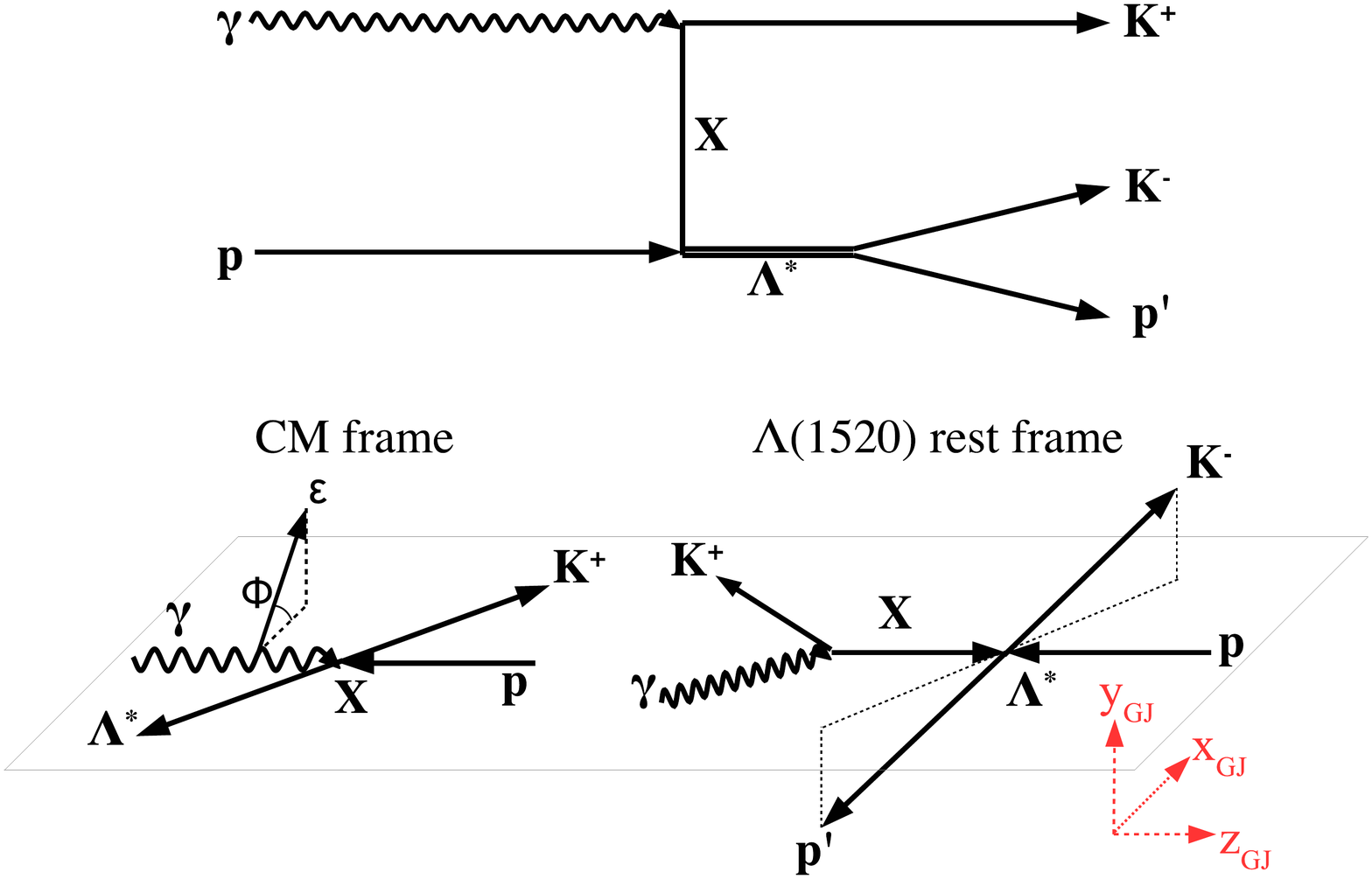}
    \end{center}
	\caption[Gottfried-Jackson system]{The Gottfried-Jackson system used in this analysis. The diagram on top visualizes the t\nobreakdash-channel production process expected to dominate at GlueX energies with $X$ being the exchange particle. The Gottfried-Jackson system is defined in the rest frame of the $\Lambda(1520)$ (see Eq.~\eqref{eq:GJ}). The polarization vector of the incoming beam photon is denoted by $\epsilon$.}
	\label{fig:formalism:GJframe}
\end{figure}

We can learn about the production mechanism of the $\Lambda(1520)$ photoproduction by studying the spin transferred to it from the polarized photon. The spin density matrix $\rho$ quantifies the spin polarization of the $\Lambda(1520)$ and parameterizes the angular distribution of its decay into $K^-p$. At high photon energies, $t$\nobreakdash-channel exchange is expected to dominate this reaction, so it is convenient to study it in the Gottfried-Jackson (GJ), or $t$\nobreakdash-channel helicity, system~\cite{Gottfried1964}. The coordinates are defined as 
\begin{align}
	\hat{z} = \frac{-\vec{p}_{p}}{\left|-\vec{p}_{p}\right|} \,, \hspace{0.5cm}
	\hat{y} = \frac{\vec{p}_{\gamma}\times\vec{p}_{K^{+}}}{\left|\vec{p}_{\gamma}\times\vec{p}_{K^{+}}\right|} \,, \hspace{0.5cm}
	\hat{x} = \hat{y}\times\hat{z} \,, \label{eq:GJ}
\end{align}
with $\vec{p}_{p/\gamma/K^{+}}$ denoting the 3-momentum of the target proton, incoming beam photon, and $K^+$ in the rest frame of the $\Lambda(1520)$. This is illustrated in Fig.~\ref{fig:formalism:GJframe}.\par
As the $\Lambda(1520)$ is a spin-\nicefrac{3}{2} particle, it has a $4\times4$ spin density matrix with 16 complex matrix elements. They are denoted by $\rho_{2\lambda_\Lambda,2\lambda'_\Lambda}$, where $\lambda_\Lambda$ denotes the $\Lambda(1520)$ helicity. Using a linearly polarized photon beam to produce the $\Lambda(1520)$ makes it possible to decompose the spin density matrix into
\begin{align}
    \rho = \rho^0-P_\gamma\cos2\Phi\rho^1-P_\gamma\sin2\Phi\rho^2 \,,
\end{align}
where $P_\gamma$ is the polarization of the photon beam and $\Phi$ is the angle between the photon polarization plane and the hadronic production plane, which is defined by the incoming $\gamma$ and target proton and the outgoing $K^+$ and $\Lambda(1520)$ (see Fig.~\ref{fig:formalism:GJframe}).
Studying the decay $\Lambda(1520)\rightarrow K^-p$ means ten SDMEs, four unpolarized and six polarized, are accessible. To measure them, the distributions of $\theta$ and $\phi$ of the $K^-$ in the GJ system are studied. These are given by Eq.~\eqref{eq:theory:SDMEfunction} below~\cite{Yu2017}. This intensity distribution is normalized in such a way that integration over angles leads to the measured differential cross section $d\sigma/dt$,  given the standard normalization $\rho^0_{33} +\rho^0_{11} = \frac{1}{2}$. There are thus nine independent SDMEs:
\begin{widetext}
    \begin{align}
    	W(\theta,\phi,\Phi)	&= \frac{1}{2\pi} \frac{d\sigma}{dt}\frac{3}{4\pi}\left\{\rho^0_{33}\sin^2{\theta}+ \rho^0_{11}\left(\frac{1}{3}+\cos^2{\theta}\right) - \frac{2}{\sqrt{3}}\text{Re}\rho^0_{31}\sin{2\theta}\cos{\phi} - \frac{2}{\sqrt{3}}\text{Re}\rho^0_{3-1}\sin^2{\theta}\cos{2\phi}\right.\nonumber\\
    	&- P_\gamma\cos{2\Phi} \left[\rho^1_{33}\sin^2{\theta}+ \rho^1_{11}\left(\frac{1}{3}+\cos^2{\theta}\right) - \frac{2}{\sqrt{3}}\text{Re}\rho^1_{31}\sin{2\theta}\cos{\phi} - \frac{2}{\sqrt{3}}\text{Re}\rho^1_{3-1}\sin^2{\theta}\cos{2\phi}\right]\nonumber\\
    	&- P_\gamma\sin{2\Phi} \frac{2}{\sqrt{3}}\left[\text{Im}\rho^2_{31}\sin{2\theta}\sin{\phi} + \text{Im}\rho^2_{3-1}\sin^2{\theta}\sin{2\phi}\right]\left.\vphantom{\frac{1}{2}}\right\}. \label{eq:theory:SDMEfunction}
    \end{align}
\end{widetext}\par
In order to relate the spin of the particle to the production exchange mechanism, Schilling~\textit{et~al.} showed that certain combinations of SDMEs can be expressed as linear combinations of purely natural or purely unnatural exchange amplitudes~\cite{Schilling1970}. The naturality for a particle with spin-parity quantum number $J^P$ is defined as $\eta=P(-1)^J$. As such, vector and tensor mesons (e.g. $K^*(892)$ and $K_2^*(1430)$) are natural exchanges ($\eta=+1$), and pseudoscalar and axial-vector mesons (e.g. $K(492)$ and $K_1(1270)$) are unnatural exchanges ($\eta=-1$). We denote production amplitudes for natural exchanges as $N$ and for unnatural exchanges as $U$. Working in the reflectivity basis with helicities $\lambda_\gamma = \pm1$, $\lambda_p = \pm\nicefrac{1}{2}$, and $\lambda_\Lambda = \pm\nicefrac{1}{2}, \pm\nicefrac{3}{2}$, and using the parity constraint results in four natural ($N_\sigma$) and four unnatural ($U_\sigma$) amplitudes, where $\sigma=\lambda_p-\lambda_\Lambda = \{-1,0,1,2\}$.
\begin{subequations}
\begin{align}
	\rho^{0}_{11} + \rho^{1}_{11} &= \frac{2}{\mathcal{N}}\left(|N_{0}|^{2}+|N_{1}|^{2}\right)\,, \label{eq:theory:nat1} \\
	\rho^{0}_{33} + \rho^{1}_{33} &= \frac{2}{\mathcal{N}}\left(|N_{-1}|^{2}+|N_{2}|^{2}\right)\,, \label{eq:theory:nat2}\\
	\text{Re}\left(\rho^{0}_{31} + \rho^{1}_{31}\right) &= \frac{2}{\mathcal{N}}\text{Re}\left(N_{-1}N_{0}^{*}-N_{2}N_{1}^{*}\right)\,, \label{eq:theory:nat3}\\
	\text{Re}\left( \rho^{0}_{3-1} + \rho^{1}_{3-1}\right) &= \frac{2}{\mathcal{N}}\text{Re}\left(N_{-1}N_{1}^{*}+N_{2}N_{0}^{*}\right)\,, \label{eq:theory:nat4} \\
	\rho^{0}_{11} - \rho^{1}_{11} &= \frac{2}{\mathcal{N}}\left(|U_{0}|^{2}+|U_{1}|^{2}\right)\,, \label{eq:theory:nat5}  \\
	\rho^{0}_{33} - \rho^{1}_{33} &= \frac{2}{\mathcal{N}}\left(|U_{-1}|^{2}+|U_{2}|^{2}\right)\,, \label{eq:theory:nat6} \\
	\text{Re} \left(\rho^{0}_{31} - \rho^{1}_{31}\right) &= 	\frac{2}{\mathcal{N}}\text{Re}\left(U_{-1}U_{0}^{*}-U_{2}U_{1}^{*}\right)\,, \label{eq:theory:nat7}
\end{align}
\begin{align}
	\text{Re} \left(\rho^{0}_{3-1} - \rho^{1}_{3-1}\right) &= \frac{2}{\mathcal{N}}\text{Re}\left(U_{-1}U_{1}^{*}+U_{2}U_{0}^{*}\right)\,. \label{eq:theory:nat8}
\end{align}
The normalization $\mathcal{N}$ is given by
\begin{align}
    \mathcal{N} = 2\left(|N_{-1}|^{2}+|N_{0}|^{2}+|N_{1}|^{2}+|N_{2}|^{2} \right.\nonumber\\
    \left.+|U_{-1}|^{2}+|U_{0}|^{2}+|U_{1}|^{2}+|U_{2}|^{2} \right)\,.\label{eq:theory:norm}
\end{align}
\end{subequations}
These combinations can be used to study the naturality of exchanged particles in a $t$\nobreakdash-channel exchange based on the extracted SDMEs. A full derivation of Eqs.~\eqref{eq:theory:nat1}-\eqref{eq:theory:nat8} is given in Appendix~\ref{app:derivation}.

\section{GlueX experiment}\label{sec:experiment}
\begin{figure}[htpb]
    \begin{center}
        \includegraphics[width=\linewidth]{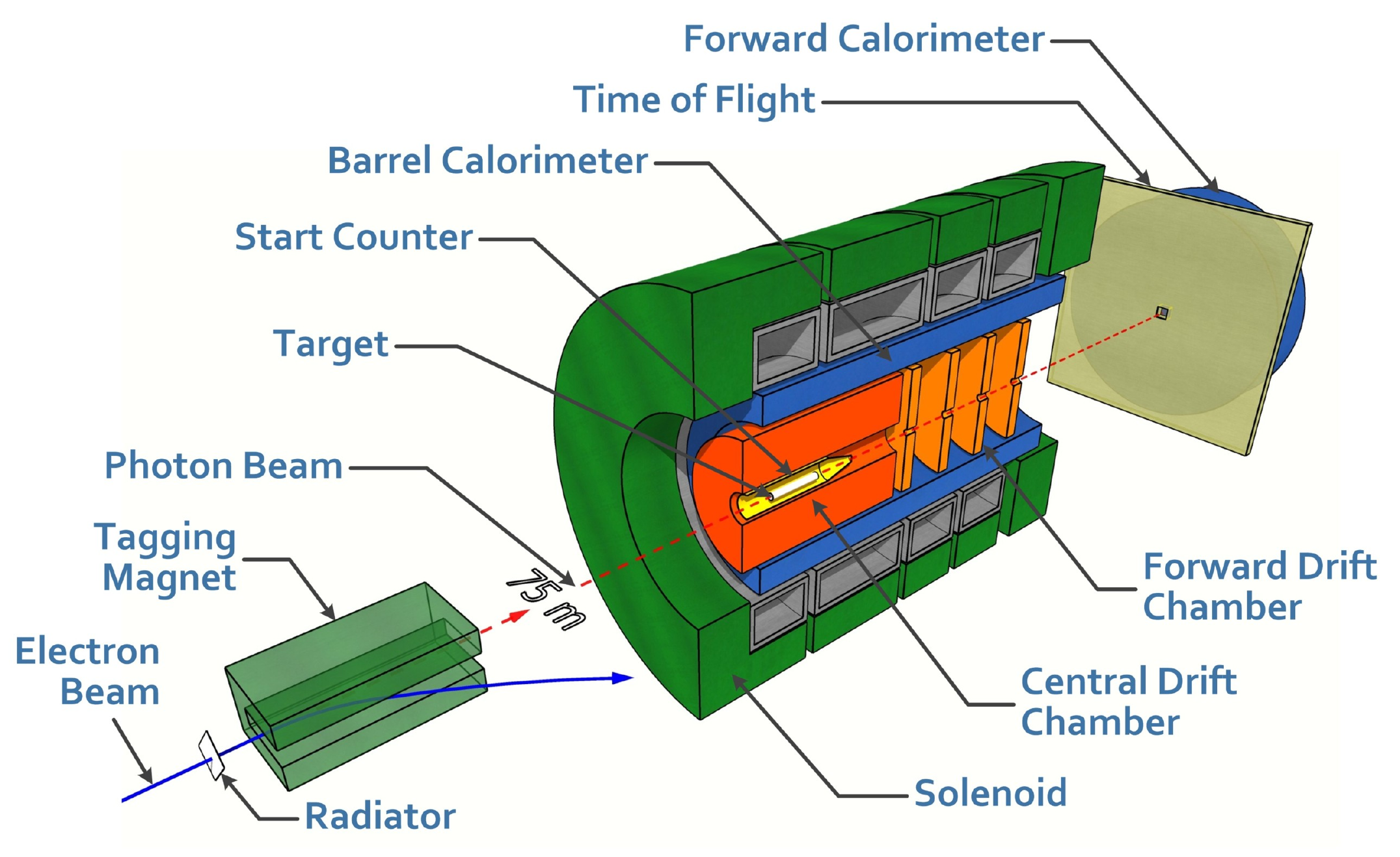}
    \end{center}
	\caption[Overview of the GlueX experiment]{Overview of the GlueX experiment and its important subdetector systems. Taken from Ref.~\cite{Adhikari2020}.}
	\label{fig:detector:GlueX}
\end{figure}

\begin{figure}
	\centering
	\includegraphics[width=0.9\linewidth]{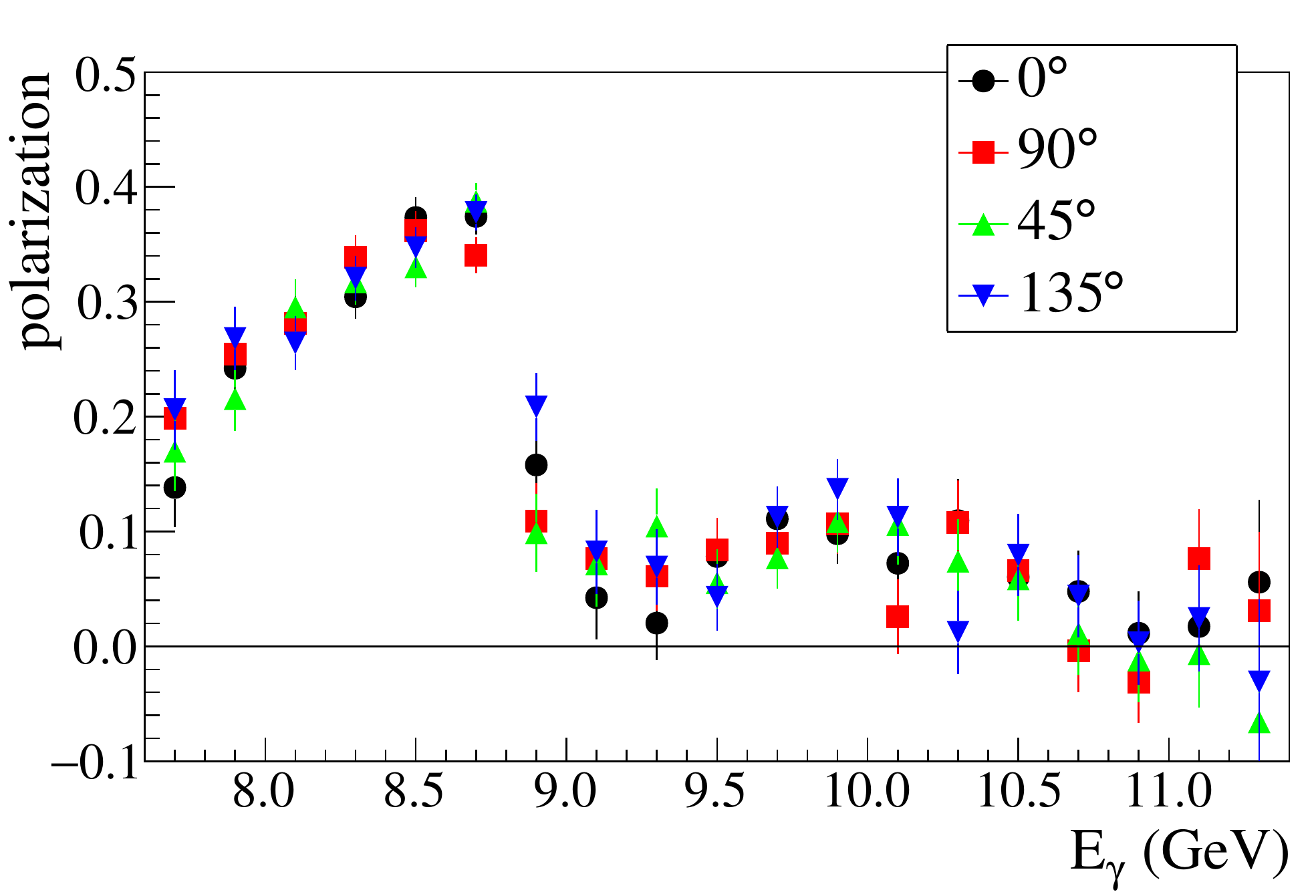}
	\caption[Photon beam polarization]{The photon beam polarization is shown for the four separate diamond settings. The measurement carries a systematic uncertainty of $1.5\%$~\cite{Dugger2017}.}
	\label{fig:detector:pol}
\end{figure}

The GlueX experiment is described in detail in Ref.~\cite{Adhikari2020}. A schematic overview is shown in Fig.~\ref{fig:detector:GlueX}. The GlueX spectrometer is located in Hall D at Jefferson Lab.
To collect the data used in this paper, an \SI{11.6}{\GeV} electron beam provided by the Continuous Electron Beam Accelerator Facility (CEBAF) was used to produce a linearly polarized photon beam via the coherent bremsstrahlung technique on a thin diamond radiator. The orientation of the beam polarization plane was controlled by adjusting the orientation of the diamond using a goniometer. During data taking, four different pairwise orthogonal diamond settings were used in turn to control systematic effects. The beam polarization had its maximum in the coherent peak, whose position was also controlled through diamond orientation. Figure~\ref{fig:detector:pol} shows the degree of polarization, as measured with a triplet polarimeter~\cite{Dugger2017}, for the four different diamond orientations. The measurement of the polarization carries a systematic uncertainty of $1.5\%$~\cite{Dugger2017}. Together with a $3\%$ statistical uncertainty, this results in an overall uncertainty on the degree of linear polarization of $\pm3.5\%$. Only events with a photon beam energy in the range from $E_\gamma=$ \SIrange{8.2}{8.8}{\GeV}, where the polarization and also the flux were highest, were analyzed. For each diamond setting, the average polarization in this range was determined and used for further analysis. In addition, for about $15\%$ of the data, an aluminum radiator was used to generate an unpolarized photon beam. Measurements using each of the beam settings (four polarized, one unpolarized) were distributed evenly across the beamtime to minimize systematic effects such as small drifts in detector acceptance or efficiency. 
\par
The electrons scattering from the radiator were deflected by a dipole magnet onto the tagger focal plane, where an array of scintillation detectors measured their position, and hence momentum, allowing the energy of associated bremsstrahlung photons to be determined. The collimated photon beam was incident on the liquid hydrogen target, which was enclosed by the Start Counter (SC), a scintillation detector that provides a reference time for each event. Surrounding this were the Central Drift Chamber (CDC) for tracking of charged particles, and the Barrel Calorimeter (BCAL), which also enclosed the Forward Drift Chamber (FDC), all in a \SI{2}{\tesla} magnetic field. The tracking detectors had a momentum resolution of $\sigma_p/p\approx1-5\%$. A Time-of-Flight detector, with the main purpose of particle identification, and the Forward Calorimeter (FCAL), were placed in the forward direction. The excellent timing resolution of the BCAL of \SI{150}{\ps} at \SI{1}{\GeV} meant that it was possible to use it for a time-of-flight measurement to help with particle identification (PID). \par
The data used in this measurement were collected in spring 2017 and correspond to an integrated luminosity of about \SI{21.8}{\pico\barn^{-1}}. The main readout trigger required a minimum energy deposition in either the BCAL or a combination of BCAL and FCAL.\par
To model the detector acceptance and reconstruction efficiency, a standardized \textsc{Geant4}-based~\cite{Agostinelli2003} GlueX detector simulation, \textit{hdgeant4}, was used~\cite{Adhikari2020}.

\section{Event selection} \label{sec:selection}
\begin{figure}
    \centering
    \includegraphics[width=0.95\linewidth]{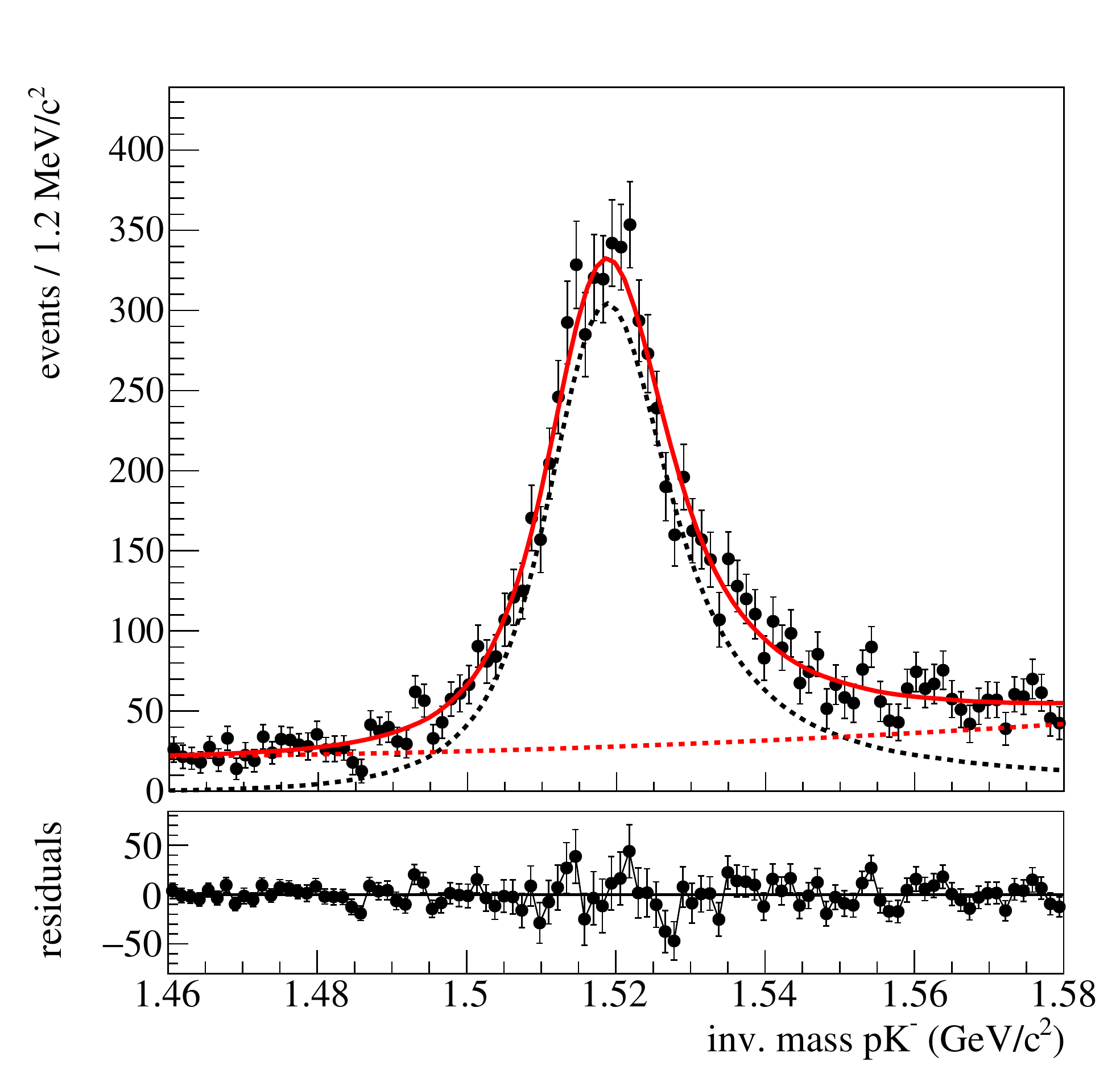}
    \caption{Example sPlot fit~\cite{Pivk2005} for one bin between $-(t-t_\text{0}) = \SIrange[range-phrase=-]{0.3}{0.5}{\GeV^{2}/\clight^2}$. Top: The black and red dashed lines show the fit components of signal and background respectively, the red solid line shows the resulting fit to the data (black points). Bottom: Residuals of the total fit to the data.}
    \label{fig:selection:splot}
\end{figure}

In order to select the reaction $\gamma p\rightarrow K^+\Lambda(1520)\rightarrow K^+K^-p$, events with at least two positively charged and one negatively charged final-state particle were analyzed. Up to three additional charged tracks were allowed to be detected in an event, to make sure that good events were not erroneously rejected because of spurious tracks in the detector. Each combination of two positive tracks and one negative track was analyzed as $K^+K^-p$.
For particle identification, time-of-flight requirements were placed for each track, using the detector with the best available timing information. Furthermore, the energy loss $dE/dx$ of the proton in the CDC was used for PID. A kinematic fit was carried out with the fit hypothesis $\gamma p\rightarrow K^+K^-p$, which included vertex and four-momentum constraints. Events with a kinematic fit confidence level of $\text{CL}<10^{-6}$ were rejected. Also, it was required that the particle tracks originated from within the target cell.
To restrict events to the $\Lambda(1520)$ signal region, only those with a $pK^-$ invariant mass between \SI{1.46}{\GeV/\clight^2} and \SI{1.58}{\GeV/\clight^2} were analyzed (see Fig.~\ref{fig:selection:splot}).\par
The electron beam, and hence the photon beam, had a bunch structure with a timing separation of \SI{4.008}{\nano\second}. Each bunch resulted in multiple hits on the tagger, of which only one belonged to the beam photon that triggered the event. This beam photon was determined via a coincidence between the hadronic event time and the bunch time. To remove the background from photons within the same bunch, a statistical sideband subtraction was performed. For this, tagger hits that were recorded close in time before and after the bunch in coincidence with the hadronic event were analyzed. These events were given a negative weight proportional to the relative size of coincidence peak and sideband regions, which were defined based on the time difference between the tagger hit and the beam bunch.\par
The sPlot technique~\cite{Pivk2005}, which was successfully used in other experiments extracting polarization observables~\cite{MAINZ-A2:2016iua,A2atMAMI:2021iuz}, was used to subtract the remaining background under the $\Lambda(1520)$ signal peak by determining event-by-event sWeights, which, when applied to the data, resulted in the signal distribution used for further analysis. This was done in eight bins of four-momentum transfer $-(t-t_\text{0})$, where $t$ is the Mandelstam variable that describes the transfer of four-momentum from the beam photon to the target proton. Its kinematic limit is given by $t_\text{0}$. To create a signal shape, events were generated according to a relativistic Breit-Wigner distribution and then simulated with \textit{hdgeant4}. Additional parameters were added to the signal function to allow flexibility in accounting for small differences between data and simulation. A second-degree Chebyshev polynomial was used to parameterize the background under the $\Lambda(1520)$. The fit was performed within the \textit{brufit} framework~\cite{Glaziera}, which uses \textsc{RooFit}~\cite{Verkerke2003}. An example fit is shown in Fig.~\ref{fig:selection:splot}. The dashed black and red lines show the signal and background contributions, respectively. The solid red line is the total fit to the data (black points). In addition, the fit residuals, which are used to assess the fit quality, are shown. The plots show that the chosen distributions describe the data very well. To test a potential impact of the background model on the results, three additional variations of Chebyshev polynomials were tested. No significant systematic effect was observed.
\par
\begin{figure}
    \centering
    \includegraphics[width=0.95\linewidth]{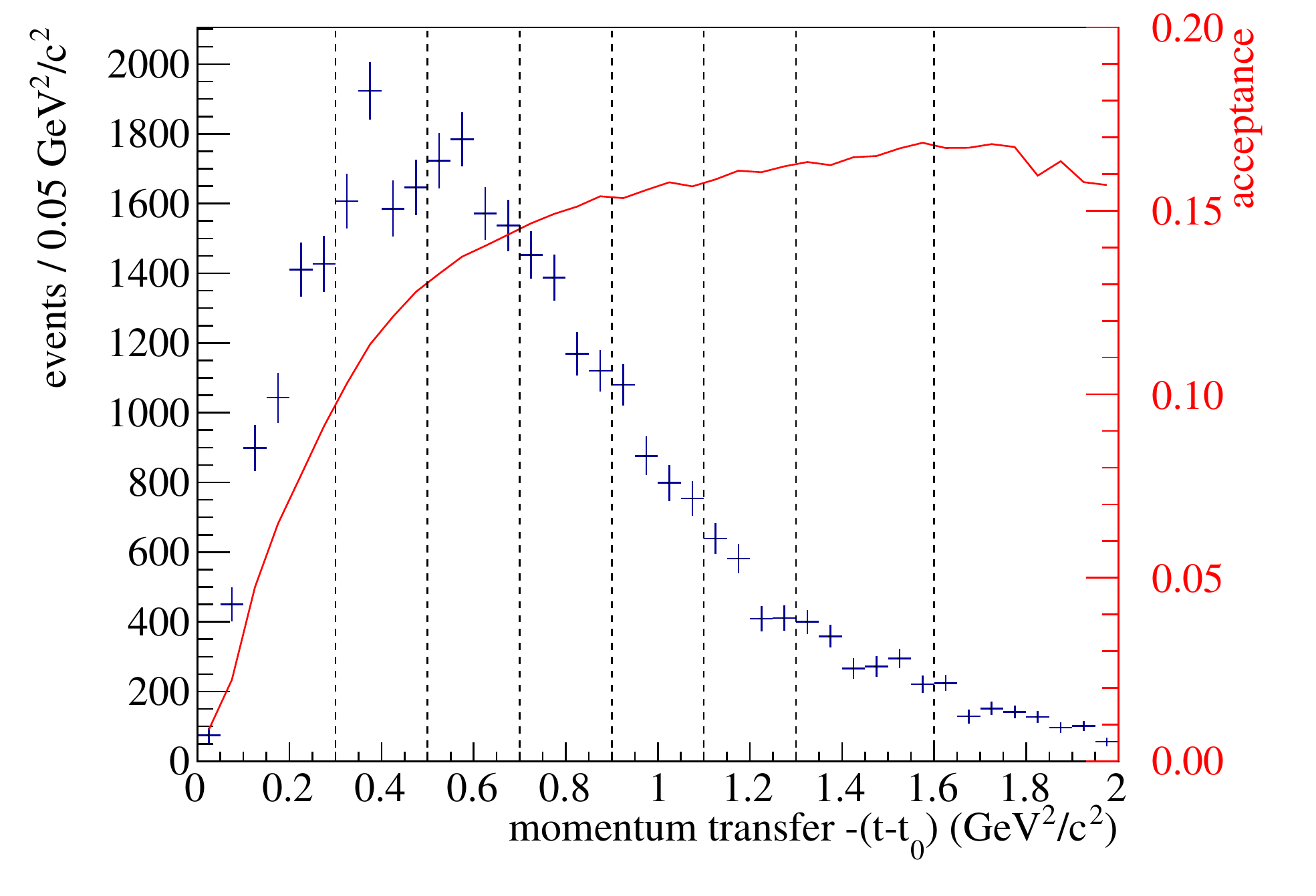}
    \caption{Distribution of momentum transfer $-(t-t_\text{0})$ for signal weighted events (blue data points) and acceptance (red line). The black dotted lines indicate the chosen bin limits.}
    \label{fig:selection:momT}
\end{figure}

After applying the sPlot background subtraction, about 32,200 events remained for the extraction of SDMEs. Their $-(t-t_\text{0})$ distribution is shown in Fig.~\ref{fig:selection:momT}. The dashed black lines indicate the bin limits used in this analysis. The solid red line represents the acceptance as determined from simulations.

\section{SDME parameter estimation} \label{sec:MCMC}
To estimate the nine independent SDME parameters, the Markov Chain Monte Carlo (MCMC) technique was used~\cite{Bonamente2017}. Instead of minimizing a $\chi^2$ or negative log-likelihood, as is often done, this method explores the possible parameter space numerically using the Metropolis-Hastings (MH) algorithm~\cite{Metropolis:1953am,HASTINGS1970}. For this purpose, a likelihood function was written as 
\begin{align}
    \ln{\mathcal{L}} = s_w\left( \sum_{i=1}^{N} {}_sw_i\ln{\mathcal{I}} - \int\text{d}\Omega\,\mathcal{I}\,\eta(\Omega) \right)\,, \label{eq:likelihood}
\end{align}
with $\mathcal{I} = W(\theta,\phi,\Phi)$ (Eq.~\eqref{eq:theory:SDMEfunction}) being the intensity function. The sWeights are notated by ${}_sw_i$, and
\begin{align}
    {s_w=\frac{\sum_{i=1}^{N}{}_sw_i}{\sum_{i=1}^{N}{}_sw_i^2}}
\end{align}
is a constant factor accounting for the effect of the weights on the statistical uncertainty. While the sum in the likelihood ran over all $N$ events in the dataset, the integral was evaluated as a sum over simulated data that were generated flat in phase space and processed through \textit{hdgeant4}. This accounted for detector acceptance effects denoted as $\eta(\Omega)$ in the likelihood function.\par
For this analysis, the MH implementation of \textsc{RooStats}~\cite{Moneta2010} was used.
As prior, a uniform distribution of SDME values with range $[-1,1]$ was assumed, reflecting the fact that SDMEs are confined to this region. New steps in the Markov chain were proposed by a \textit{sequential proposal} function which randomly changed one of the nine SDME parameters at a time and proposed its next value based on a Gaussian distribution centered around the current value with a width tuned to achieve an acceptance rate of about $10-20\%$.
Final parameters are reported as the means of the posterior distributions, with uncertainties given by the widths. \par
\begin{figure}
	\centering
	\includegraphics[width=0.49\linewidth]{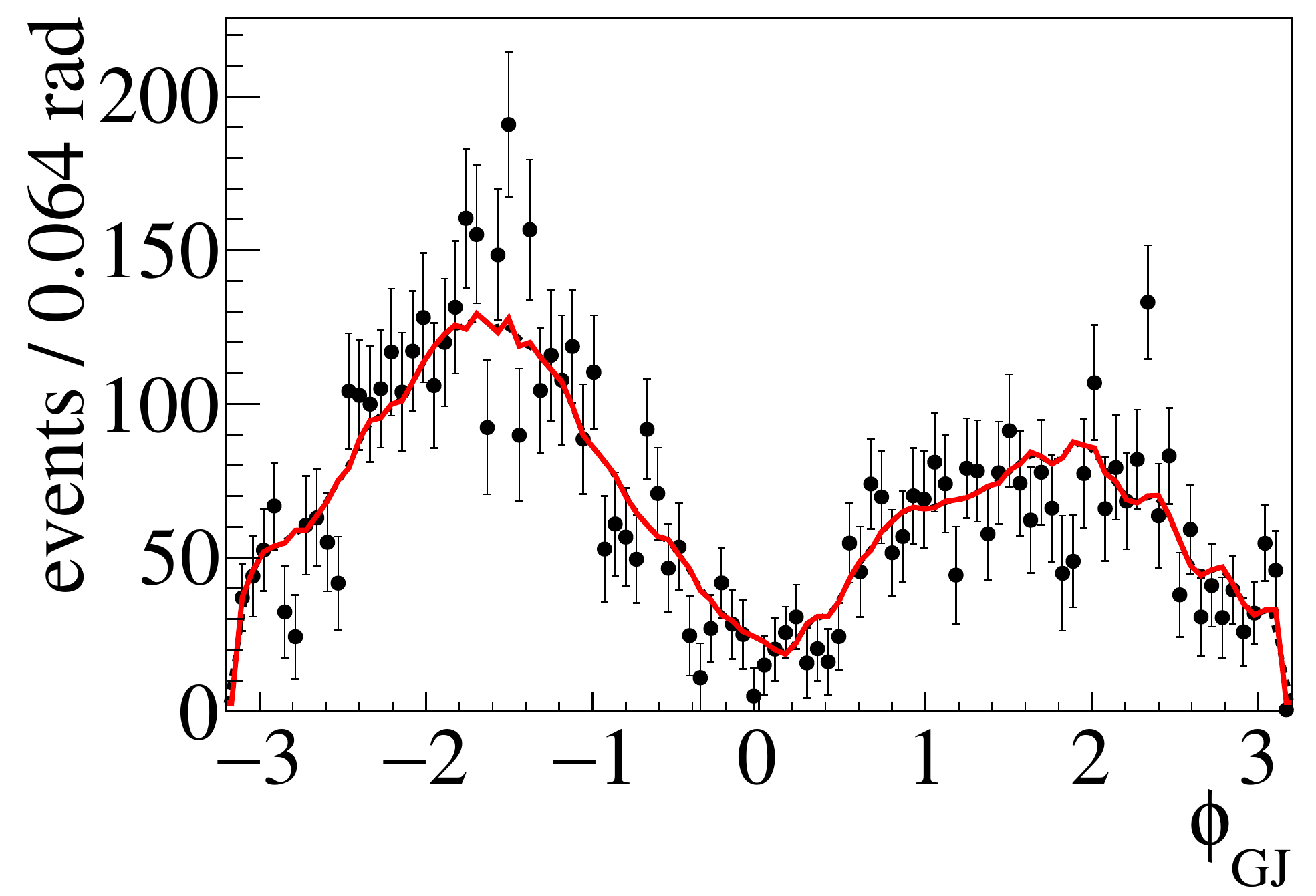}
	\includegraphics[width=0.49\linewidth]{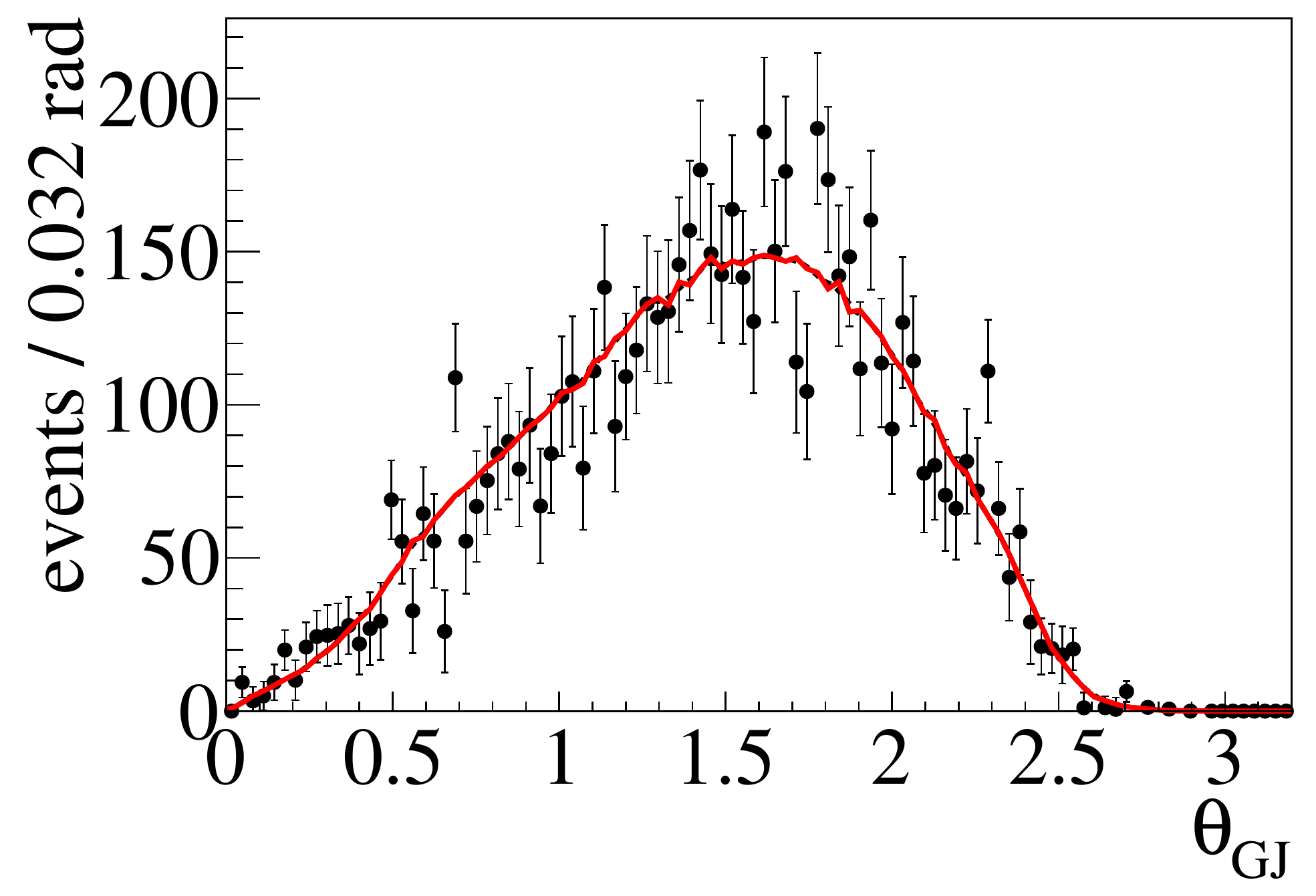}
	\caption{Projections of data (black points) and simulations containing acceptance effects which were weighted with fit results (red lines) for $-(t-t_\text{0}) = \SIrange[range-phrase=-]{0.3}{0.5}{\GeV^{2}/\clight^2}$. Note, as the red curves are based on simulated data, they are not expected to be perfectly smooth.}
	\label{fig:selection:sdmeFits4}
\end{figure}

In order to visually assess the quality of the extracted SDME parameters, simulations that were produced flat in phase space and processed through \textit{hdgeant4} to incorporate detector inefficiencies were re-weighted with the resulting intensity function. A comparison of this weighted simulation to data, for one example bin and the two variables $\theta_{GJ}$ and $\phi_{GJ}$, is shown in Fig.~\ref{fig:selection:sdmeFits4}.

\subsection{Validation}
Extensive studies on simulated data have been performed to validate that the presented approach to extract the SDME parameters gave on average the correct results and uncertainties for the estimated parameters. Each study included simulations of 400 statistically independent samples of signal and background with pre-selected, known SDMEs. The simulations were processed through the complete \textit{hdgeant4} simulation and treated as real data from that point onwards, including the full event selection and sPlot background subtraction. For each sample, the SDMEs were extracted and compared to the parameters chosen for generation. Observed differences were quantified in a systematic uncertainty (see next section). Details on all studies can be found in Ref.~\cite{Pauli2020a}.

\subsection{Systematic uncertainties}
As described earlier, the relative systematic uncertainty on the beam polarization was determined to be $\pm3.5\%$. This only affected the polarized $\rho^{1,2}$ SDMEs and the uncertainties are fully correlated across the full $-(t-t_0)$ range. The absolute systematic uncertainty of the extraction method of SDME parameters, obtained from the validation studies showed only very little correlation across the $-(t-t_0)$ range and was $\pm0.02$ for unpolarized and $\pm0.01$ for polarized SDMEs. In addition to these and the assumed background distribution in the sPlot fit, other aspects of the analysis were tested carefully for systematic effects. Twenty-six different variations in the event selection, including different limits for the $pK^-$ invariant mass range, the kinematic fit confidence level, vertex position, and timing of the particles in the BCAL and TOF were considered. None of them showed significant systematic effects on the results. The full list of tested variations can be found in Ref.~\cite{Pauli2020a}.\par
In order to explore systematic uncertainties, due to inaccuracies in the simulated model of the angular dependence of the tracking efficiency, we used the results of a study similar to that described in Sec.~15.1 of Ref.~\cite{Adhikari2020}. For each of the three tracks in the event we obtained the ratio of efficiency in data to efficiency in simulation for the particular region in the two-dimensional momentum\nobreakdash-$\theta$ plane. For tracks beyond the region of phase space covered in Ref.~\cite{Adhikari2020}, we used their largest measured ratio. We then reweighted the accepted simulated data by the product of these weights for each of the three tracks and repeated the SDME analysis. We observed that the absolute central values of the SDMEs changed by no more than $\pm0.01$ for the unpolarized $\rho^0$ and $\pm0.007$ for the polarized $\rho^{1,2}$, with little correlation across the $-(t-t_0)$ range. We therefore used this as our estimate for the systematic uncertainty due to inaccuracies in the simulated model and combine it with the other systematic uncertainties in quadrature.\par
A summary of the relevant systematic uncertainties is given in Table~\ref{tab:systematics}. The total systematic uncertainty for each individual bin is given in Table~\ref{tab:numerical:results} together with the results.
\begin{table}[]
\centering
\caption{Summary of systematic uncertainties. The uncertainties for extraction method and simulation model are absolute numbers while the uncertainty on the degree of polarization is a relative scaling uncertainty.}
\label{tab:systematics}
\begin{ruledtabular}
\begin{tabular}{l|cc}
source & uncertainty $\rho^0$ & uncertainty $\rho^{1,2}$  \\ \hline
extraction method      & $0.02$ & $0.01$  \\ 
simulation model       & $0.01$ & $0.007$ \\ 
degree of polarization & -    & $3.5\%$ 
\end{tabular} 
\end{ruledtabular}
\end{table}

\section{Results} \label{sec:results}
\begin{figure*}[htbp]
    \includegraphics[width=0.85\linewidth]{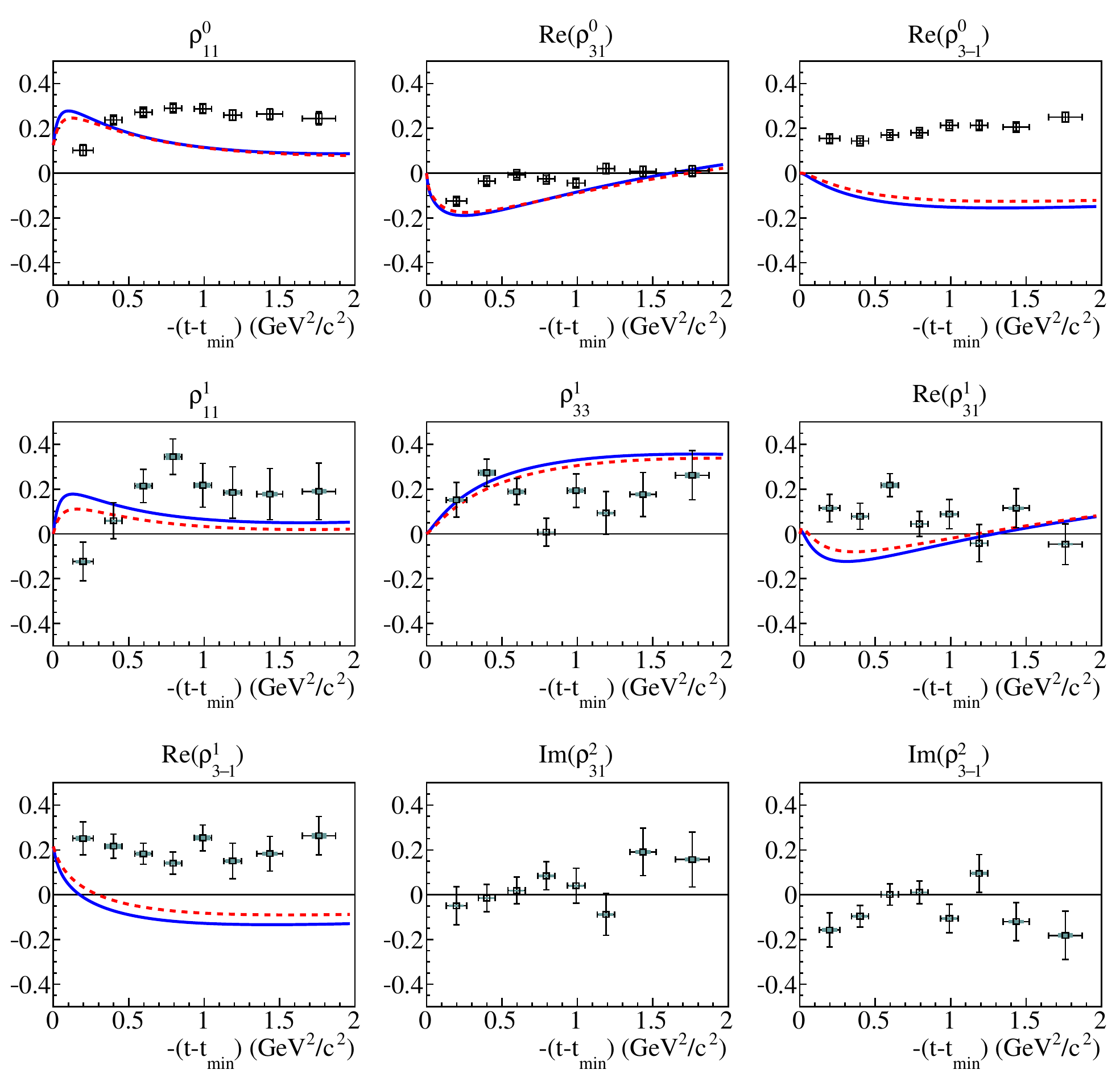}
    \caption{Spin density matrix elements and predictions by Yu and Kong (based on Ref.~\cite{Yu2017}), using parameters based on data from CLAS~\cite{Moriya2013} and LEPS~\cite{Muramatsu2009,Kohri2010} (blue solid) and using parameters based on data from LAMP2~\cite{Barber1980} and SLAC~\cite{Boyarski1971} (red dashed). The vertical error bars show the statistical uncertainty, the blue shaded boxes the scaling uncertainty from the polarization, and the black boxes the remaining systematic uncertainties combined in quadrature. The horizontal error bars show the RMS widths within the $-(t-t_0)$ bins.}
    \label{fig:results:sdmes}
\end{figure*}

Results are shown in Fig.~\ref{fig:results:sdmes}. The vertical error bars show the statistical uncertainty, the blue shaded boxes the scaling uncertainty from the polarization, and the black box the remaining systematic uncertainties combined in quadrature. The horizontal error bars show the RMS widths within the $-(t-t_0)$ bins. Also shown in the figure are predictions made by Yu and Kong (private communication based on Ref.~\cite{Yu2017}) for seven of the nine extracted SDMEs. The blue solid lines show the predictions based on data from CLAS~\cite{Moriya2013} and LEPS~\cite{Muramatsu2009,Kohri2010}, and the red dashed lines show predictions based on data from LAMP2~\cite{Barber1980} and SLAC~\cite{Boyarski1971}. These predictions are based on previous data at much lower or higher photon beam energies, and until now there have been no data for polarized SDMEs. It is clear that these new GlueX data will place stringent new constraints on the model. \par
\begin{figure}[htbp]
    \includegraphics[width=\linewidth]{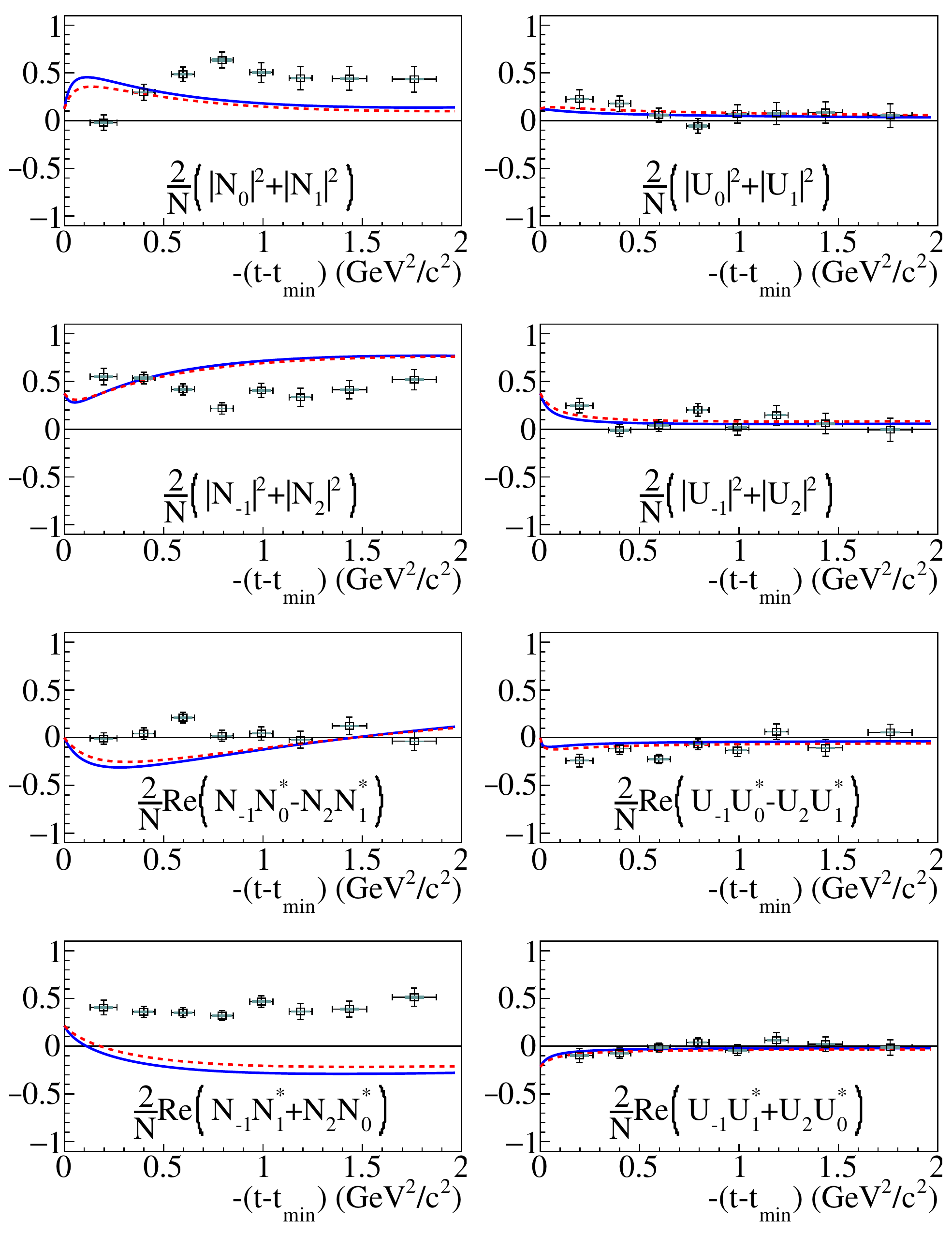}
    \caption{Combinations of SDMEs corresponding to natural (left column) and unnatural (right column) exchanges in the photoproduction of $\Lambda(1520)$. Also shown are the corresponding combinations of predictions by Yu and Kong (based on Ref.~\cite{Yu2017}), using parameters based on data from CLAS~\cite{Moriya2013} and LEPS~\cite{Muramatsu2009,Kohri2010} (blue solid) and using parameters based on data from LAMP2~\cite{Barber1980} and SLAC~\cite{Boyarski1971} (red dashed). The vertical error bars show the statistical uncertainty, the blue shaded boxes the scaling uncertainty from the polarization, and the black boxes the remaining systematic uncertainties combined in quadrature. The horizontal error bars show the RMS widths within the $-(t-t_0)$ bins.}
    \label{fig:results:naturality}
\end{figure}

To interpret the extracted SDMEs in terms of the contributing exchange mechanism, the combinations from Eqs.~\eqref{eq:theory:nat1}-\eqref{eq:theory:nat8} were formed and are shown in Fig.~\ref{fig:results:naturality}.
Over most of the $-(t-t_\text{0})$ range, the results indicate natural exchanges are dominant. Only in the lowest bin, there seems to be a small contribution from unnatural exchanges. Although the observed dominance of natural exchange does not allow us to specify a particular exchange particle, we note that Yu and Kong predicted a dominant natural exchange at high energies, via a $K_2^*$. This can be seen by looking at the same combinations for their predictions. The unnatural contributions are 0 over most of the $-(t-t_\text{0})$ range. While their model does not agree well with the present SDME measurements, the expectation of a large natural contribution to the exchange is supported by the data.

\section{Conclusion} \label{sec:conclusion}
Nine independent spin density matrix elements for the reaction $\gamma p\rightarrow K^+\Lambda(1520)\rightarrow K^+K^-p$ have been measured for photon beam energies between \SIrange{8.2}{8.8}{\GeV}. For the $\Lambda(1520)$ this represents the first measurement of polarized SDMEs and the first measurement of unpolarized SDMEs  at these energies. Our measurements allow conclusions about the production mechanisms by studying combinations of SDMEs that represent purely natural or unnatural exchanges in a $t$\nobreakdash-channel exchange. It was found that the photoproduction of $\Lambda(1520)$ at these energies is dominated by natural exchange amplitudes over most of the $-(t-t_\text{0})$ range, in qualitative agreement with the only available model prediction. However, the quantitative agreement with the model was poor. The presented results will hopefully motivate more work on the photoproduction of $\Lambda(1520)$ which should lead to better agreement between data and models in the future. To further the understanding of this reaction, precise measurements of differential cross sections are desirable. GlueX is ideally placed to perform these over a wide range of energies.

\begin{acknowledgments}
We would like to acknowledge the outstanding efforts of the staff of the Accelerator and the Physics Divisions at Jefferson Lab that made the experiment possible. We also would like to thank B.-G. Yu for helpful discussions and sharing his calculations with us. This work was supported in part by the U.S. Department of Energy, the U.S. National Science Foundation, the German Research Foundation, Forschungszentrum J\"ulich GmbH, GSI Helmholtzzentrum f\"ur Schwerionenforschung GmbH, the Natural Sciences and Engineering Research Council of Canada, the Russian Foundation for Basic Research, the UK Science and Technology Facilities Council, the Chilean Comisi\'{o}n Nacional de Investigaci\'{o}n Cient\'{i}fica y Tecnol\'{o}gica, the National Natural Science Foundation of China and the China Scholarship Council. V. Mathieu is a Serra H\'unter fellow and acknowledges support from the Community of Madrid through 
the Programa de atracción de talento investigador 2018-T1/TIC-10313 and from the Spanish national grant PID2019-106080GB-C21. This material is based upon work supported by the U.S. Department of Energy, Office of Science, Office of Nuclear Physics under contract DE-AC05-06OR23177.
\end{acknowledgments}

\appendix
\section{Derivation of natural and unnatural amplitudes} \label{app:derivation}
We provide a brief derivation of Eqs.~\eqref{eq:theory:nat1}-\eqref{eq:theory:nat8}, which are used to interpret the SDMEs in terms of the naturality $\eta=P(-1)^J$ of the exchanged particle. \par
Following Schilling~\cite{Schilling1970}, we denote the production amplitude as $T_{\lambda_\gamma\lambda_p\lambda_\Lambda}$ with helicities $\lambda_\gamma = \pm1$, $\lambda_p=\pm\nicefrac{1}{2}$, and $\lambda_\Lambda=\pm\nicefrac{1}{2},\pm\nicefrac{3}{2}$. Taking parity $T_{-\lambda_\gamma-\lambda_p-\lambda_\Lambda}=(-1)^{\lambda_p-\lambda_\Lambda}T_{\lambda_\gamma\lambda_p\lambda_\Lambda}$ into account, this leaves us with eight independent amplitudes. We can split each amplitude into a positive (N, $\epsilon=+1$) and a negative (U, $\epsilon=-1$) component and write
\begin{align}
    T = T^{(+)} + T^{(-)}\,,
\end{align}
with the amplitudes given in reflectivity basis by
\begin{align}
    T^{(\epsilon)}_{\lambda_p\lambda_\Lambda} = \frac{1}{2}\left( T_{1\lambda_p\lambda_\Lambda}+\epsilon T_{-1\lambda_p\lambda_\Lambda}\right)\,.
\end{align}
We can use these amplitudes to express the SDMEs as
\begin{align}
	\rho^{0}_{\lambda_{\Lambda}\lambda_{\Lambda}'} = & \frac{1}{\mathcal{N}} \sum_{\lambda_{p}}T^{(+)}_{\lambda_{p}\lambda_{\Lambda}} T^{(+)*}_{\lambda_{p}\lambda_{\Lambda}'} + T^{(-)}_{\lambda_{p}\lambda_{\Lambda}} T^{(-)*}_{\lambda_{p}\lambda_{\Lambda}'}\,, \\
	\rho^{1}_{\lambda_{\Lambda}\lambda_{\Lambda}'} = & \frac{1}{\mathcal{N}} \sum_{\lambda_{p}}T^{(+)}_{\lambda_{p}\lambda_{\Lambda}} T^{(+)*}_{\lambda_{p}\lambda_{\Lambda}'} - T^{(-)}_{\lambda_{p}\lambda_{\Lambda}} T^{(-)*}_{\lambda_{p}\lambda_{\Lambda}'}\,, \\
	\rho^{2}_{\lambda_{\Lambda}\lambda_{\Lambda}'} = & \frac{i}{\mathcal{N}} \sum_{\lambda_{p}}T^{(+)}_{\lambda_{p}\lambda_{\Lambda}} T^{(-)*}_{\lambda_{p}\lambda_{\Lambda}'} - T^{(-)}_{\lambda_{p}\lambda_{\Lambda}} T^{(+)*}_{\lambda_{p}\lambda_{\Lambda}'}\,,
\end{align}
and write the following combinations
\begin{align}
	\rho^{0}_{\lambda_{\Lambda}\lambda_{\Lambda}'}+\rho^{1}_{\lambda_{\Lambda}\lambda_{\Lambda}'}= & \frac{2}{\mathcal{N}} \sum_{\lambda_{p}}T^{(+)}_{\lambda_{p}\lambda_{\Lambda}} T^{(+)*}_{\lambda_{p}\lambda_{\Lambda}'} \label{eq:app:poscomb}\,,\\
	\rho^{0}_{\lambda_{\Lambda}\lambda_{\Lambda}'}-\rho^{1}_{\lambda_{\Lambda}\lambda_{\Lambda}'}= & \frac{2}{\mathcal{N}} \sum_{\lambda_{p}}T^{(-)}_{\lambda_{p}\lambda_{\Lambda}} T^{(-)*}_{\lambda_{p}\lambda_{\Lambda}'} \label{eq:app:negcomb}\,.
\end{align}
These combinations separate the amplitudes with positive and negative reflectivity. We can further write the amplitudes as $N_{\sigma}$ (natural) and $U_{\sigma}$ (unnatural) with $\sigma=\lambda_{p}-\lambda_{\Lambda}=\{-1,0,1,2\}$ and write the reflectivity amplitudes in terms of the exchange naturality
\begin{align}
	N_{-1} &= T_{\frac{1}{2}\frac{3}{2}}^{(+)} &N_{0} &= T_{\frac{1}{2}\frac{1}{2}}^{(+)} & N_{1} &=T_{\frac{1}{2}-\frac{1}{2}}^{(+)} & N_{2} &=T_{\frac{1}{2}-\frac{3}{2}}^{(+)} \label{eq:app:nat} \\
	U_{-1} &= T_{\frac{1}{2}\frac{3}{2}}^{(-)} &U_{0} &= T_{\frac{1}{2}\frac{1}{2}}^{(-)} & U_{1} &=T_{\frac{1}{2}-\frac{1}{2}}^{(-)} & U_{2} &=T_{\frac{1}{2}-\frac{3}{2}}^{(-)} \label{eq:app:unnat}
\end{align}
Using Eqs.~\eqref{eq:app:nat} and \eqref{eq:app:unnat}, with Eqs.~\eqref{eq:app:poscomb} and \eqref{eq:app:negcomb}, leads directly to Eqs.~\eqref{eq:theory:nat1}-\eqref{eq:theory:nat8} with normalization given by Eq.~\eqref{eq:theory:norm}.
\FloatBarrier

\section{Numerical results}
All numerical results for the SDMEs and their statistical and systematic uncertainties, together with the natural and unnatural combinations are listed in Table~\ref{tab:numerical:results}. In general, the correlations in the statistical uncertainties are small, except for $\rho^{1}_{11}$ and $\rho^{1}_{33}$, whose covariances need to be taken into account when using the data further, and are listed as well. \par
Subsets of the Markov chains used for the parameter estimation are available as supplemental material~\cite{suppl}.
\begingroup
\onecolumngrid
\squeezetable
\begin{turnpage}
\begin{table*}[htbp]
	\caption{Numerical results for all presented SDMEs, natural and unnatural combinations, and covariances between $\rho^{1}_{11}$ and $\rho^{1}_{33}$. The first uncertainty is statistical, the second systematic}
	\label{tab:numerical:results}
	\begin{ruledtabular}
	\begin{tabular}{c|ccccccccc}
        \begin{tabular}[c]{@{}c@{}}$-(t-t_\text{0})$\\ in $\si{GeV^{2}/\clight^{2}}$\end{tabular} & $\rho^0_{11}$ & $\rho^0_{31}$& $\rho^0_{3-1}$ & $\rho^1_{11}$ & $\rho^1_{33}$ & $\rho^1_{31}$ & $\rho^1_{3-1}$ & $\rho^2_{31}$ & $\rho^2_{3-1}$  \\ \hline
        $0.197\pm0.069$ &\begin{tabular}[c]{@{}cc@{}}\phantom{-}0.102&$\pm$0.025\\&$\pm$0.022\end{tabular} &\begin{tabular}[c]{@{}cc@{}}-0.125&$\pm$0.016\\&$\pm$0.022\end{tabular} &\begin{tabular}[c]{@{}cc@{}}\phantom{-}0.154&$\pm$0.018\\&$\pm$0.022\end{tabular} &\begin{tabular}[c]{@{}cc@{}}-0.123&$\pm$0.087\\&$\pm$0.013\end{tabular} &\begin{tabular}[c]{@{}cc@{}}\phantom{-}0.152&$\pm$0.078\\&$\pm$0.013\end{tabular} &\begin{tabular}[c]{@{}cc@{}}\phantom{-}0.115&$\pm$0.061\\&$\pm$0.013\end{tabular} &\begin{tabular}[c]{@{}cc@{}}\phantom{-}0.251&$\pm$0.074\\&$\pm$0.015\end{tabular} &\begin{tabular}[c]{@{}cc@{}}-0.049&$\pm$0.085\\&$\pm$0.012\end{tabular} &\begin{tabular}[c]{@{}cc@{}}-0.158&$\pm$0.076\\&$\pm$0.013\end{tabular} \\ 
        $0.400\pm0.056$ &\begin{tabular}[c]{@{}cc@{}}\phantom{-}0.238&$\pm$0.016\\&$\pm$0.022\end{tabular} &\begin{tabular}[c]{@{}cc@{}}-0.036&$\pm$0.015\\&$\pm$0.022\end{tabular} &\begin{tabular}[c]{@{}cc@{}}\phantom{-}0.144&$\pm$0.013\\&$\pm$0.022\end{tabular} &\begin{tabular}[c]{@{}cc@{}}\phantom{-}0.059&$\pm$0.081\\&$\pm$0.012\end{tabular} &\begin{tabular}[c]{@{}cc@{}}\phantom{-}0.273&$\pm$0.061\\&$\pm$0.016\end{tabular} &\begin{tabular}[c]{@{}cc@{}}\phantom{-}0.078&$\pm$0.059\\&$\pm$0.013\end{tabular} &\begin{tabular}[c]{@{}cc@{}}\phantom{-}0.217&$\pm$0.054\\&$\pm$0.014\end{tabular} &\begin{tabular}[c]{@{}cc@{}}-0.015&$\pm$0.060\\&$\pm$0.012\end{tabular} &\begin{tabular}[c]{@{}cc@{}}-0.096&$\pm$0.049\\&$\pm$0.013\end{tabular} \\ 
        $0.597\pm0.057$ &\begin{tabular}[c]{@{}cc@{}}\phantom{-}0.272&$\pm$0.015\\&$\pm$0.022\end{tabular} &\begin{tabular}[c]{@{}cc@{}}-0.007&$\pm$0.013\\&$\pm$0.022\end{tabular} &\begin{tabular}[c]{@{}cc@{}}\phantom{-}0.170&$\pm$0.012\\&$\pm$0.022\end{tabular} &\begin{tabular}[c]{@{}cc@{}}\phantom{-}0.214&$\pm$0.074\\&$\pm$0.014\end{tabular} &\begin{tabular}[c]{@{}cc@{}}\phantom{-}0.189&$\pm$0.058\\&$\pm$0.014\end{tabular} &\begin{tabular}[c]{@{}cc@{}}\phantom{-}0.218&$\pm$0.051\\&$\pm$0.014\end{tabular} &\begin{tabular}[c]{@{}cc@{}}\phantom{-}0.183&$\pm$0.047\\&$\pm$0.014\end{tabular} &\begin{tabular}[c]{@{}cc@{}}\phantom{-}0.019&$\pm$0.060\\&$\pm$0.012\end{tabular} &\begin{tabular}[c]{@{}cc@{}}\phantom{-}0.001&$\pm$0.048\\&$\pm$0.012\end{tabular} \\ 
        $0.793\pm0.057$ &\begin{tabular}[c]{@{}cc@{}}\phantom{-}0.290&$\pm$0.016\\&$\pm$0.022\end{tabular} &\begin{tabular}[c]{@{}cc@{}}-0.025&$\pm$0.013\\&$\pm$0.022\end{tabular} &\begin{tabular}[c]{@{}cc@{}}\phantom{-}0.180&$\pm$0.012\\&$\pm$0.022\end{tabular} &\begin{tabular}[c]{@{}cc@{}}\phantom{-}0.345&$\pm$0.079\\&$\pm$0.017\end{tabular} &\begin{tabular}[c]{@{}cc@{}}\phantom{-}0.008&$\pm$0.062\\&$\pm$0.012\end{tabular} &\begin{tabular}[c]{@{}cc@{}}\phantom{-}0.045&$\pm$0.056\\&$\pm$0.012\end{tabular} &\begin{tabular}[c]{@{}cc@{}}\phantom{-}0.141&$\pm$0.050\\&$\pm$0.013\end{tabular} &\begin{tabular}[c]{@{}cc@{}}\phantom{-}0.084&$\pm$0.063\\&$\pm$0.013\end{tabular} &\begin{tabular}[c]{@{}cc@{}}\phantom{-}0.011&$\pm$0.051\\&$\pm$0.012\end{tabular} \\ 
        $0.992\pm0.058$ &\begin{tabular}[c]{@{}cc@{}}\phantom{-}0.288&$\pm$0.019\\&$\pm$0.022\end{tabular} &\begin{tabular}[c]{@{}cc@{}}-0.044&$\pm$0.017\\&$\pm$0.022\end{tabular} &\begin{tabular}[c]{@{}cc@{}}\phantom{-}0.213&$\pm$0.014\\&$\pm$0.022\end{tabular} &\begin{tabular}[c]{@{}cc@{}}\phantom{-}0.217&$\pm$0.098\\&$\pm$0.014\end{tabular} &\begin{tabular}[c]{@{}cc@{}}\phantom{-}0.193&$\pm$0.075\\&$\pm$0.014\end{tabular} &\begin{tabular}[c]{@{}cc@{}}\phantom{-}0.088&$\pm$0.065\\&$\pm$0.013\end{tabular} &\begin{tabular}[c]{@{}cc@{}}\phantom{-}0.254&$\pm$0.058\\&$\pm$0.015\end{tabular} &\begin{tabular}[c]{@{}cc@{}}\phantom{-}0.040&$\pm$0.078\\&$\pm$0.012\end{tabular} &\begin{tabular}[c]{@{}cc@{}}-0.106&$\pm$0.064\\&$\pm$0.013\end{tabular} \\ 
        $1.189\pm0.058$ &\begin{tabular}[c]{@{}cc@{}}\phantom{-}0.259&$\pm$0.024\\&$\pm$0.022\end{tabular} &\begin{tabular}[c]{@{}cc@{}}\phantom{-}0.021&$\pm$0.021\\&$\pm$0.022\end{tabular} &\begin{tabular}[c]{@{}cc@{}}\phantom{-}0.214&$\pm$0.018\\&$\pm$0.022\end{tabular} &\begin{tabular}[c]{@{}cc@{}}\phantom{-}0.185&$\pm$0.115\\&$\pm$0.014\end{tabular} &\begin{tabular}[c]{@{}cc@{}}\phantom{-}0.094&$\pm$0.096\\&$\pm$0.013\end{tabular} &\begin{tabular}[c]{@{}cc@{}}-0.042&$\pm$0.083\\&$\pm$0.012\end{tabular} &\begin{tabular}[c]{@{}cc@{}}\phantom{-}0.150&$\pm$0.080\\&$\pm$0.013\end{tabular} &\begin{tabular}[c]{@{}cc@{}}-0.088&$\pm$0.094\\&$\pm$0.013\end{tabular} &\begin{tabular}[c]{@{}cc@{}}\phantom{-}0.095&$\pm$0.084\\&$\pm$0.013\end{tabular} \\ 
        $1.435\pm0.086$ &\begin{tabular}[c]{@{}cc@{}}\phantom{-}0.264&$\pm$0.025\\&$\pm$0.022\end{tabular} &\begin{tabular}[c]{@{}cc@{}}\phantom{-}0.008&$\pm$0.024\\&$\pm$0.022\end{tabular} &\begin{tabular}[c]{@{}cc@{}}\phantom{-}0.206&$\pm$0.019\\&$\pm$0.022\end{tabular} &\begin{tabular}[c]{@{}cc@{}}\phantom{-}0.178&$\pm$0.114\\&$\pm$0.014\end{tabular} &\begin{tabular}[c]{@{}cc@{}}\phantom{-}0.176&$\pm$0.098\\&$\pm$0.014\end{tabular} &\begin{tabular}[c]{@{}cc@{}}\phantom{-}0.115&$\pm$0.087\\&$\pm$0.013\end{tabular} &\begin{tabular}[c]{@{}cc@{}}\phantom{-}0.183&$\pm$0.077\\&$\pm$0.014\end{tabular} &\begin{tabular}[c]{@{}cc@{}}\phantom{-}0.191&$\pm$0.106\\&$\pm$0.014\end{tabular} &\begin{tabular}[c]{@{}cc@{}}-0.120&$\pm$0.085\\&$\pm$0.013\end{tabular} \\ 
        $1.761\pm0.111$ &\begin{tabular}[c]{@{}cc@{}}\phantom{-}0.244&$\pm$0.030\\&$\pm$0.022\end{tabular} &\begin{tabular}[c]{@{}cc@{}}\phantom{-}0.010&$\pm$0.024\\&$\pm$0.022\end{tabular} &\begin{tabular}[c]{@{}cc@{}}\phantom{-}0.250&$\pm$0.021\\&$\pm$0.022\end{tabular} &\begin{tabular}[c]{@{}cc@{}}\phantom{-}0.190&$\pm$0.127\\&$\pm$0.014\end{tabular} &\begin{tabular}[c]{@{}cc@{}}\phantom{-}0.262&$\pm$0.110\\&$\pm$0.015\end{tabular} &\begin{tabular}[c]{@{}cc@{}}-0.046&$\pm$0.091\\&$\pm$0.012\end{tabular} &\begin{tabular}[c]{@{}cc@{}}\phantom{-}0.263&$\pm$0.086\\&$\pm$0.015\end{tabular} &\begin{tabular}[c]{@{}cc@{}}\phantom{-}0.157&$\pm$0.122\\&$\pm$0.013\end{tabular} &\begin{tabular}[c]{@{}cc@{}}-0.182&$\pm$0.108\\&$\pm$0.014\end{tabular} \\ \hline
        \hline
        \begin{tabular}[c]{@{}c@{}}$-(t-t_\text{0})$\\ in $\si{GeV^{2}/\clight^{2}}$\end{tabular} &
        $\sigma_{\rho^{1}_{11}\rho^{1}_{33}}$ &
        $\frac{2}{\mathcal{N}}\left(|N_{0}|^{2}+|N_{1}|^{2}\right)$ &
        $\frac{2}{\mathcal{N}}\left(|N_{-1}|^{2}+|N_{2}|^{2}\right)$ &
        \begin{tabular}[c]{l@{}r@{}}$\frac{2}{\mathcal{N}}\text{Re}\left(N_{-1}N_{0}^{*}\right.$\\\hspace{1cm}$\left.-N_{2}N_{1}^{*}\right)$\end{tabular} &
        \begin{tabular}[c]{l@{}r@{}}$\frac{2}{\mathcal{N}}\text{Re}\left(N_{-1}N_{1}^{*}\right.$\\\hspace{1cm}$\left.+N_{2}N_{0}^{*}\right)$\end{tabular} &
        $\frac{2}{\mathcal{N}}\left(|U_{0}|^{2}+|U_{1}|^{2}\right)$ &
        $\frac{2}{\mathcal{N}}\left(|U_{-1}|^{2}+|U_{2}|^{2}\right)$ &
        \begin{tabular}[c]{l@{}r@{}}$\frac{2}{\mathcal{N}}\text{Re}\left(U_{-1}U_{0}^{*}\right.$\\\hspace{1cm}$\left.-U_{2}U_{1}^{*}\right)$\end{tabular} &
        \begin{tabular}[c]{l@{}r@{}}$\frac{2}{\mathcal{N}}\text{Re}\left(U_{-1}U_{1}^{*}\right.$\\\hspace{1cm}$\left.+U_{2}U_{0}^{*}\right)$\end{tabular} \\ \hline
        $0.197\pm0.069$ &-0.00438& \begin{tabular}[c]{@{}cc@{}}-0.021&$\pm$0.081\\&$\pm$0.035\end{tabular} &\begin{tabular}[c]{@{}cc@{}}\phantom{-}0.550&$\pm$0.087\\&$\pm$0.035\end{tabular} &\begin{tabular}[c]{@{}cc@{}}-0.010&$\pm$0.060\\&$\pm$0.035\end{tabular} &\begin{tabular}[c]{@{}cc@{}}\phantom{-}0.405&$\pm$0.077\\&$\pm$0.036\end{tabular} &\begin{tabular}[c]{@{}cc@{}}\phantom{-}0.225&$\pm$0.099\\&$\pm$0.035\end{tabular} &\begin{tabular}[c]{@{}cc@{}}\phantom{-}0.246&$\pm$0.076\\&$\pm$0.035\end{tabular} &\begin{tabular}[c]{@{}cc@{}}-0.240&$\pm$0.066\\&$\pm$0.035\end{tabular} &\begin{tabular}[c]{@{}cc@{}}-0.097&$\pm$0.075\\&$\pm$0.036\end{tabular} \\ 
        $0.400\pm0.056$ &-0.00368& \begin{tabular}[c]{@{}cc@{}}\phantom{-}0.296&$\pm$0.085\\&$\pm$0.035\end{tabular} &\begin{tabular}[c]{@{}cc@{}}\phantom{-}0.536&$\pm$0.060\\&$\pm$0.036\end{tabular} &\begin{tabular}[c]{@{}cc@{}}\phantom{-}0.043&$\pm$0.061\\&$\pm$0.035\end{tabular} &\begin{tabular}[c]{@{}cc@{}}\phantom{-}0.360&$\pm$0.057\\&$\pm$0.035\end{tabular} &\begin{tabular}[c]{@{}cc@{}}\phantom{-}0.179&$\pm$0.080\\&$\pm$0.035\end{tabular} &\begin{tabular}[c]{@{}cc@{}}-0.011&$\pm$0.066\\&$\pm$0.036\end{tabular} &\begin{tabular}[c]{@{}cc@{}}-0.114&$\pm$0.059\\&$\pm$0.035\end{tabular} &\begin{tabular}[c]{@{}cc@{}}-0.073&$\pm$0.053\\&$\pm$0.035\end{tabular} \\ 
        $0.597\pm0.057$ &-0.00321& \begin{tabular}[c]{@{}cc@{}}\phantom{-}0.486&$\pm$0.077\\&$\pm$0.035\end{tabular} &\begin{tabular}[c]{@{}cc@{}}\phantom{-}0.417&$\pm$0.058\\&$\pm$0.035\end{tabular} &\begin{tabular}[c]{@{}cc@{}}\phantom{-}0.210&$\pm$0.057\\&$\pm$0.035\end{tabular} &\begin{tabular}[c]{@{}cc@{}}\phantom{-}0.353&$\pm$0.051\\&$\pm$0.035\end{tabular} &\begin{tabular}[c]{@{}cc@{}}\phantom{-}0.058&$\pm$0.074\\&$\pm$0.035\end{tabular} &\begin{tabular}[c]{@{}cc@{}}\phantom{-}0.039&$\pm$0.062\\&$\pm$0.035\end{tabular} &\begin{tabular}[c]{@{}cc@{}}-0.225&$\pm$0.050\\&$\pm$0.035\end{tabular} &\begin{tabular}[c]{@{}cc@{}}-0.013&$\pm$0.046\\&$\pm$0.035\end{tabular} \\ 
        $0.793\pm0.057$ &-0.00363& \begin{tabular}[c]{@{}cc@{}}\phantom{-}0.635&$\pm$0.084\\&$\pm$0.037\end{tabular} &\begin{tabular}[c]{@{}cc@{}}\phantom{-}0.218&$\pm$0.061\\&$\pm$0.034\end{tabular} &\begin{tabular}[c]{@{}cc@{}}\phantom{-}0.019&$\pm$0.059\\&$\pm$0.035\end{tabular} &\begin{tabular}[c]{@{}cc@{}}\phantom{-}0.321&$\pm$0.053\\&$\pm$0.035\end{tabular} &\begin{tabular}[c]{@{}cc@{}}-0.055&$\pm$0.077\\&$\pm$0.037\end{tabular} &\begin{tabular}[c]{@{}cc@{}}\phantom{-}0.202&$\pm$0.068\\&$\pm$0.034\end{tabular} &\begin{tabular}[c]{@{}cc@{}}-0.070&$\pm$0.057\\&$\pm$0.035\end{tabular} &\begin{tabular}[c]{@{}cc@{}}\phantom{-}0.039&$\pm$0.050\\&$\pm$0.035\end{tabular} \\ 
        $0.992\pm0.058$ &-0.00544& \begin{tabular}[c]{@{}cc@{}}\phantom{-}0.505&$\pm$0.102\\&$\pm$0.035\end{tabular} &\begin{tabular}[c]{@{}cc@{}}\phantom{-}0.406&$\pm$0.075\\&$\pm$0.035\end{tabular} &\begin{tabular}[c]{@{}cc@{}}\phantom{-}0.044&$\pm$0.070\\&$\pm$0.035\end{tabular} &\begin{tabular}[c]{@{}cc@{}}\phantom{-}0.467&$\pm$0.062\\&$\pm$0.036\end{tabular} &\begin{tabular}[c]{@{}cc@{}}\phantom{-}0.070&$\pm$0.097\\&$\pm$0.035\end{tabular} &\begin{tabular}[c]{@{}cc@{}}\phantom{-}0.019&$\pm$0.080\\&$\pm$0.035\end{tabular} &\begin{tabular}[c]{@{}cc@{}}-0.133&$\pm$0.065\\&$\pm$0.035\end{tabular} &\begin{tabular}[c]{@{}cc@{}}-0.041&$\pm$0.057\\&$\pm$0.036\end{tabular} \\ 
        $1.189\pm0.058$ &-0.00772& \begin{tabular}[c]{@{}cc@{}}\phantom{-}0.444&$\pm$0.120\\&$\pm$0.035\end{tabular} &\begin{tabular}[c]{@{}cc@{}}\phantom{-}0.334&$\pm$0.096\\&$\pm$0.035\end{tabular} &\begin{tabular}[c]{@{}cc@{}}-0.021&$\pm$0.089\\&$\pm$0.035\end{tabular} &\begin{tabular}[c]{@{}cc@{}}\phantom{-}0.364&$\pm$0.084\\&$\pm$0.035\end{tabular} &\begin{tabular}[c]{@{}cc@{}}\phantom{-}0.074&$\pm$0.115\\&$\pm$0.035\end{tabular} &\begin{tabular}[c]{@{}cc@{}}\phantom{-}0.147&$\pm$0.102\\&$\pm$0.035\end{tabular} &\begin{tabular}[c]{@{}cc@{}}\phantom{-}0.062&$\pm$0.083\\&$\pm$0.035\end{tabular} &\begin{tabular}[c]{@{}cc@{}}\phantom{-}0.063&$\pm$0.080\\&$\pm$0.035\end{tabular} \\ 
        $1.435\pm0.086$ &-0.00794& \begin{tabular}[c]{@{}cc@{}}\phantom{-}0.441&$\pm$0.123\\&$\pm$0.035\end{tabular} &\begin{tabular}[c]{@{}cc@{}}\phantom{-}0.412&$\pm$0.095\\&$\pm$0.035\end{tabular} &\begin{tabular}[c]{@{}cc@{}}\phantom{-}0.124&$\pm$0.092\\&$\pm$0.035\end{tabular} &\begin{tabular}[c]{@{}cc@{}}\phantom{-}0.389&$\pm$0.082\\&$\pm$0.035\end{tabular} &\begin{tabular}[c]{@{}cc@{}}\phantom{-}0.086&$\pm$0.111\\&$\pm$0.035\end{tabular} &\begin{tabular}[c]{@{}cc@{}}\phantom{-}0.060&$\pm$0.107\\&$\pm$0.035\end{tabular} &\begin{tabular}[c]{@{}cc@{}}-0.107&$\pm$0.089\\&$\pm$0.035\end{tabular} &\begin{tabular}[c]{@{}cc@{}}\phantom{-}0.022&$\pm$0.077\\&$\pm$0.035\end{tabular} \\ 
        $1.761\pm0.111$ &-0.00941& \begin{tabular}[c]{@{}cc@{}}\phantom{-}0.434&$\pm$0.136\\&$\pm$0.035\end{tabular} &\begin{tabular}[c]{@{}cc@{}}\phantom{-}0.518&$\pm$0.106\\&$\pm$0.036\end{tabular} &\begin{tabular}[c]{@{}cc@{}}-0.036&$\pm$0.100\\&$\pm$0.035\end{tabular} &\begin{tabular}[c]{@{}cc@{}}\phantom{-}0.513&$\pm$0.095\\&$\pm$0.036\end{tabular} &\begin{tabular}[c]{@{}cc@{}}\phantom{-}0.054&$\pm$0.124\\&$\pm$0.035\end{tabular} &\begin{tabular}[c]{@{}cc@{}}-0.006&$\pm$0.121\\&$\pm$0.036\end{tabular} &\begin{tabular}[c]{@{}cc@{}}\phantom{-}0.056&$\pm$0.087\\&$\pm$0.035\end{tabular} &\begin{tabular}[c]{@{}cc@{}}-0.013&$\pm$0.082\\&$\pm$0.036\end{tabular} 
    \end{tabular}
    \end{ruledtabular}
\end{table*}
\end{turnpage}
\twocolumngrid
\endgroup

\FloatBarrier
\bibliographystyle{apsrev4-2.bst}
\bibliography{paperref}

\end{document}